\documentclass[fleqn,usenatbib]{mnras}

\usepackage[T1]{fontenc}

\DeclareRobustCommand{\VAN}[3]{#2}
\let\VANthebibliography\thebibliography
\def\thebibliography{\DeclareRobustCommand{\VAN}[3]{##3}\VANthebibliography}

\usepackage{graphicx}	% Including figure files
\usepackage{amsmath}	% Advanced maths commands
\usepackage{amssymb}	% Extra maths symbols
\usepackage{multirow}
\usepackage{caption}
\usepackage{multicol}        % Multi-column entries in tables
\usepackage{longtable}
\usepackage{xr}

\hypersetup{bookmarksnumbered=true}
\usepackage{newtxtext,newtxmath}
\title[AT2022cmc]{Day-timescale variability in the radio light curve of the Tidal Disruption Event AT2022cmc: confirmation of a highly relativistic outflow}
\author[L. Rhodes et al.]{L Rhodes,$^{1}$\thanks{E-mail: lauren.rhodes@physics.ox.ac.uk}
J. S. Bright,$^{1}$
R. Fender$^{1,2}$
I. Sfaradi$^{3}$
D. A. Green$^{4}$
A. Horesh$^{3}$
K. Mooley$^{5,6}$
D. Pasham$^{7}$
\newauthor
S. Smartt$^{1}$
D. J. Titterington$^{4}$
A. J. van der Horst$^{9}$
D. R. A. Williams$^{10}$
\\
$^{1}$Astrophysics, The University of Oxford, Keble Road, Oxford, OX1 3RH, UK\\
$^{2}$Department of Astrophysics, University of Cape Town, Private Bag X3, Rondebosch, Cape Town, 7701, South Africa\\
$^{3}$Racah Institute of Physics, The Hebrew University of Jerusalem, Jerusalem, 91904, Israel\\
$^{4}$Cavendish Laboratory, The University of Cambridge,19 J.J. Thomson Avenue, Cambridge, CB3 0HE, UK\\
$^{5}$National Radio Astronomy Observatory, Socorro, 87801, New Mexico, USA\\
$^{6}$Cahill Center for Astronomy and Astrophysics, California Institute of Technology, Pasadena, 91125, CA, USA\\
$^{7}$Kavli Institute for Astrophysics and Space Research, Massachusetts Institute of Technology, Cambridge, 02139, MA, USA\\
$^{1,8}$ Astrophysics Research Centre, School of Mathematics and Physics, Queen's University Belfast, Belfast, BT7 1NN, UK\\
$^{9}$Department of Physics, The George Washington University, 725 21\textsuperscript{st} Street NW, Washington DC, 20052, USA\\
$^{10}$Jodrell Bank Centre for Astrophysics, School of Physics and Astronomy, The University of Manchester, Manchester, M13 9PL, UK
}
\date{Accepted XXX. Received YYY; in original form ZZZ}

\pubyear{2015}

\begin{document}

\label{firstpage}
\pagerange{\pageref{firstpage}--\pageref{lastpage}}
\maketitle

\begin{abstract}
Tidal disruption events (TDEs) are transient, multi-wavelength events in which a star is ripped apart by a supermassive black hole. Observations show that in a small fraction of TDEs, a short-lived, synchrotron emitting jet is produced. We observed the newly discovered TDE AT2022cmc with a slew of radio facilities over the first 100 days after its discovery. The light curve from the AMI--LA radio interferometer shows day-timescale variability which we attribute to a high brightness temperature emitting region as opposed to scintillation. We measure a brightness temperature of 2$\times$10\textsuperscript{15}\,K, which is unphysical for synchrotron radiation. We suggest that the measured high brightness temperature is a result of relativistic beaming caused by a jet being launched at velocities close to the speed of light along our line of sight. We infer from day-timescale variability that the jet associated with AT2022cmc has a relativistic Doppler factor of at least 16, which corresponds to a bulk Lorentz factor of at least 8 if we are observing the jet directly on axis. Such an inference is the first conclusive evidence that the radio emission observed from some TDEs is from relativistic jets because it does not rely on an outflow model. We also compare the first 100 days of radio evolution of AT2022cmc with that of the previous bright relativistic TDE, Swift J1644, and find a remarkable similarity in their evolution.
\end{abstract}


%%%%%%%%%%%%%%%%%%%%%%%%%%%%%%%%%%%%%%%%%%%%%%%%%%

%%%%%%%%%%%%%%%%% BODY OF PAPER %%%%%%%%%%%%%%%%%%

\section{Introduction}\label{intro}

Multiple wide-field radio surveys have demonstrated that up to 30\% of the radio sky is variable when observing at centimeter wavelengths \citep{2006MNRAS.370.1556B, 2011ApJ...740...65O, 2011MNRAS.412..634B, 2013PASA...30....6M, 2017MNRAS.466.1944M, 2021ApJ...923...31S}.
%Multiple wide-field radio surveys observing between 1 and 5\,GHz have demonstrated that about 2\% of the radio sky is variable on a range of timescales \citep{2011ApJ...740...65O, 2011MNRAS.412..634B, 2013PASA...30....6M,  2016ApJ...818..105M, 2017MNRAS.466.1944M, 2021ApJ...923...31S}. At higher frequencies, the percentage of variable sources increases to over 10\% \citep{2006MNRAS.370.1556B, 2013A&A...553A.107C}. 
Some variability is extrinsic, a result of \textit{scintillation}, the scattering of radio waves by free electrons in the interstellar medium resulting in random flux modulations. Variability can also be intrinsic, a result of physical changes within the emitting region that can give clues to the nature of the source. A good example of this is observed in radio data sets of blazars (persistently accreting supermassive black holes (SMBHs) that launch outflows pointing towards Earth) which are highly variable and very luminous \citep{1997ARA&A..35..445U, 2009ApJ...703..802M}. These characteristics are indicative of a highly relativistic outflow with a compact emitting region.

Blazars are not the only systems to produce jets, and not all jets are persistent: many stellar-mass black hole systems produce transient jets e.g. gamma-ray bursts or X-ray binaries \citep{2020MNRAS.496.3326R, 2005Ap&SS.300....1F, 2020NatAs...4..697B}. In the most energetic systems, the kinetic energy required for such an outflow is so high that the radiation must originate from a highly collimated jet as opposed to a spherical outflow \citep{1993Natur.361..236M, 1999ApJ...519L..17S, 2018Natur.561..355M}. The presence of a relativistic jet may be inferred by the detection of superluminal motion/expansion or the presence of a high brightness temperature component as a result of Doppler boosting (also known as relativistic beaming). Doppler boosting occurs when outflowing material is moving at velocities close to the speed of light (i.e. with high bulk Lorentz factors) and close to the observer's line of sight making it appear more luminous than in the rest frame of the material.

If a star passes too close to a SMBH, the tidal forces of the SMBH can overcome the self-gravity keeping the star together and pull it apart creating a tidal disruption event \citep[TDE;][]{1988Natur.333..523R}. Approximately half of the disrupted material is thought to be lost while the other half falls back and is accreted onto the SMBH. In a small fraction of TDEs, the
%TDEs are currently split into two categories: those discovered due to the presence of optical/ultra-violet counterparts called \textit{thermal TDEs} and those where the 
radiation at all wavelengths is dominated by luminous non-thermal emission thought to be produced by a jet. These systems are called \textit{relativistic TDEs}. The most well-studied relativistic TDE to date is \textit{Swift} J1644+57 which was first discovered due to a bright gamma-ray flash followed by a luminous, variable and decaying X-ray counterpart \citep{2011Sci...333..203B, 2011Natur.476..421B}. A bright radio counterpart of \textit{Swift} J1644+57 has been detected at low frequencies for the past decade and modelled as a narrow, highly relativistic jet pointed towards Earth \citep{2011Sci...333..203B, 2012ApJ...748...36B, 2013ApJ...767..152Z, 2014ApJ...788...32W, 2015MNRAS.450.2824M, 2018ApJ...854...86E}. The presence of a jet would be directly confirmed upon either the detection of superluminal motion/expansion \citep{2016MNRAS.462L..66Y} or the presence of a high brightness temperature as a result of Doppler boosting, but no such evidence has yet been found.

ZTF22aaajecp is a panchromatic transient discovered by the Zwicky Transient Facility on 2022 Feb 11 10:42 UT \citep[MJD 59621.4458, $T_{0}$, ][]{2022TNSAN..38....1A} and registered on the Transient Name Server as AT2022cmc. The optical light curve shows a fast decay ($>$1\,mag/day) until day 10 post-discovery when it settled into a plateau between 21 and 22\,magnitudes \citep[r-band, ][]{2022GCN.31846....1P, 2022GCN.31805....1D, 2022GCN.31729....1C, 2022TNSAN..40....1F}. Early optical spectroscopy detected interstellar medium absorption from a likely host galaxy at $z = 1.193$ \citep{2022GCN.31602....1T}. AT2022cmc is over four times more distant than \textit{Swift} J1644+57 \citep[assuming $H_0$ = 70\,km\,s\textsuperscript{-1}\,Mpc\textsuperscript{-1} and $\Omega_{\textrm{M}}$ = 0.3, ][]{2011Sci...333..199L}. Rapid radio, sub-mm, and X-ray follow-up observations were also performed and bright counterparts were detected in all bands \citep[e.g.][]{2022GCN.31627....1P, 2022ATel15269....1A, 2022GCN.31665....1D, 2022GCN.31641....1Y, 2022GCN.31601....1P}. The high X-ray luminosity, hour-timescale variability and spectra from NICER and NuSTAR \citep{2022ATel15349....1H, 2022ATel15230....1Y, 2022ATel15232....1P} indicated that AT2022cmc is a relativistic TDE, the first to be discovered in over a decade \citep{2011Sci...333..203B, 2012ApJ...753...77C, 2015MNRAS.452.4297B}.

%Swift, 

\begin{figure*}
    \centering
    \includegraphics[width = 0.85\textwidth]{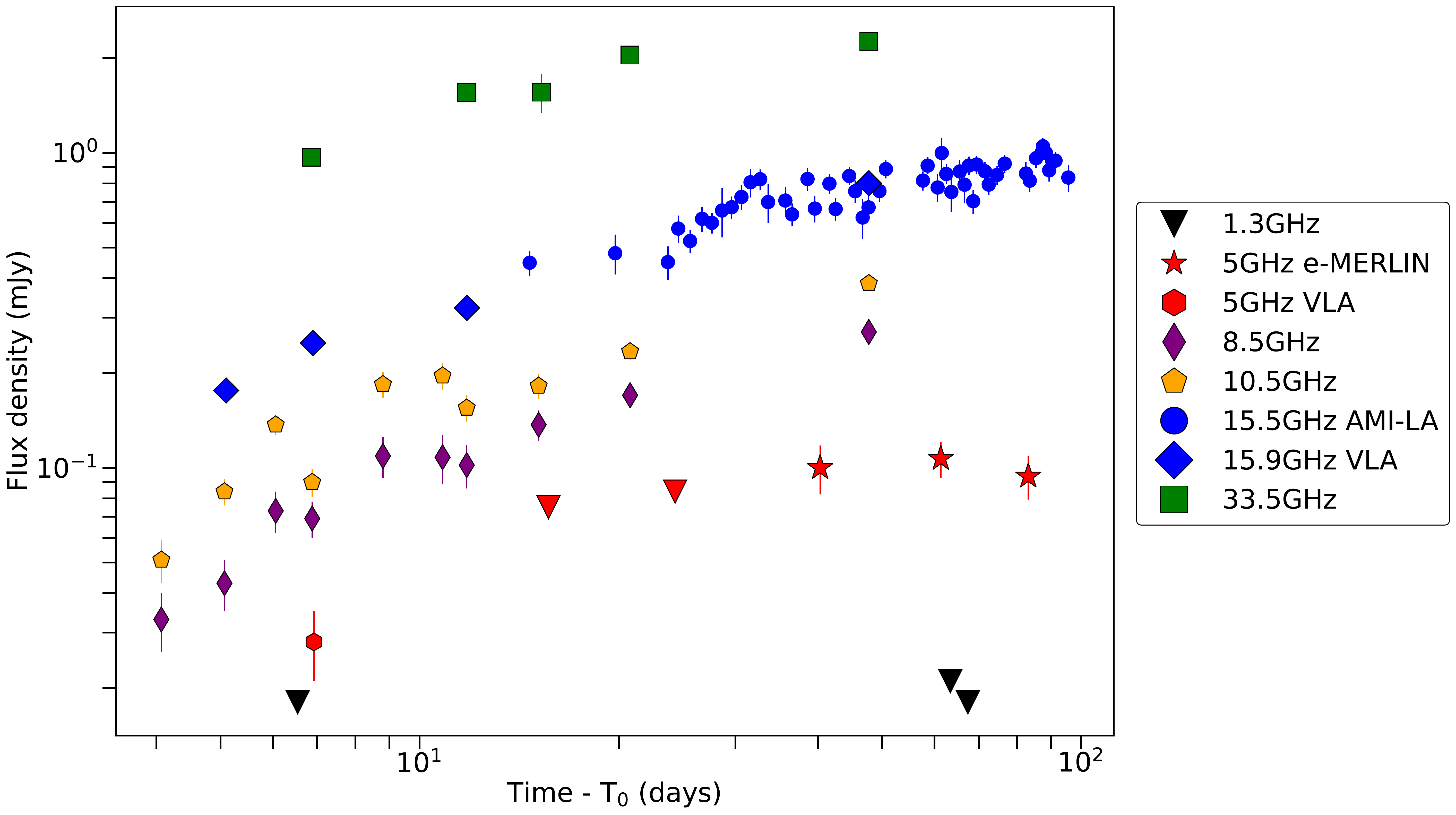}
    \caption{The radio light curve of AT2022cmc, combining the observations presented in this work at 3$\sigma$ upper limits at 1.3\,GHz (black downwards facing triangles), \textit{e}--MERLIN upper limits and detections at 5\,GHz as red downwards facing triangles and stars, respectively) and 15.5\,GHz detections and upper limits shown as blue circles and downwards facing triangles, respectively. Also shown are detections at 5 (red hexagons), 8.5 (purple narrow diamonds), 10 (gold pentagons), 15.9 (blue wide diamonds) and 33.5\,GHz (green squares) from the VLA \citep{2022arXiv221116530A}. The 15.5\,GHz light curve shows clear evidence of inter-observation variability. There is also evidence of variability at 10.5\,GHz.}
    \label{fig:lc}
\end{figure*}

\section{Methods}\label{methods}

\subsection{Observations}

We obtained radio observations of AT2022cmc through guaranteed and rapid response time on the Arcminute Microkelvin Imager Large Array (AMI-LA), \textit{enhanced} -- Multi-Element Radio Linked Interferometer Network (\textit{e}--MERLIN) and MeerKAT radio telescopes.

\begin{figure*}
    \centering
    \includegraphics[width = 0.7\textwidth]{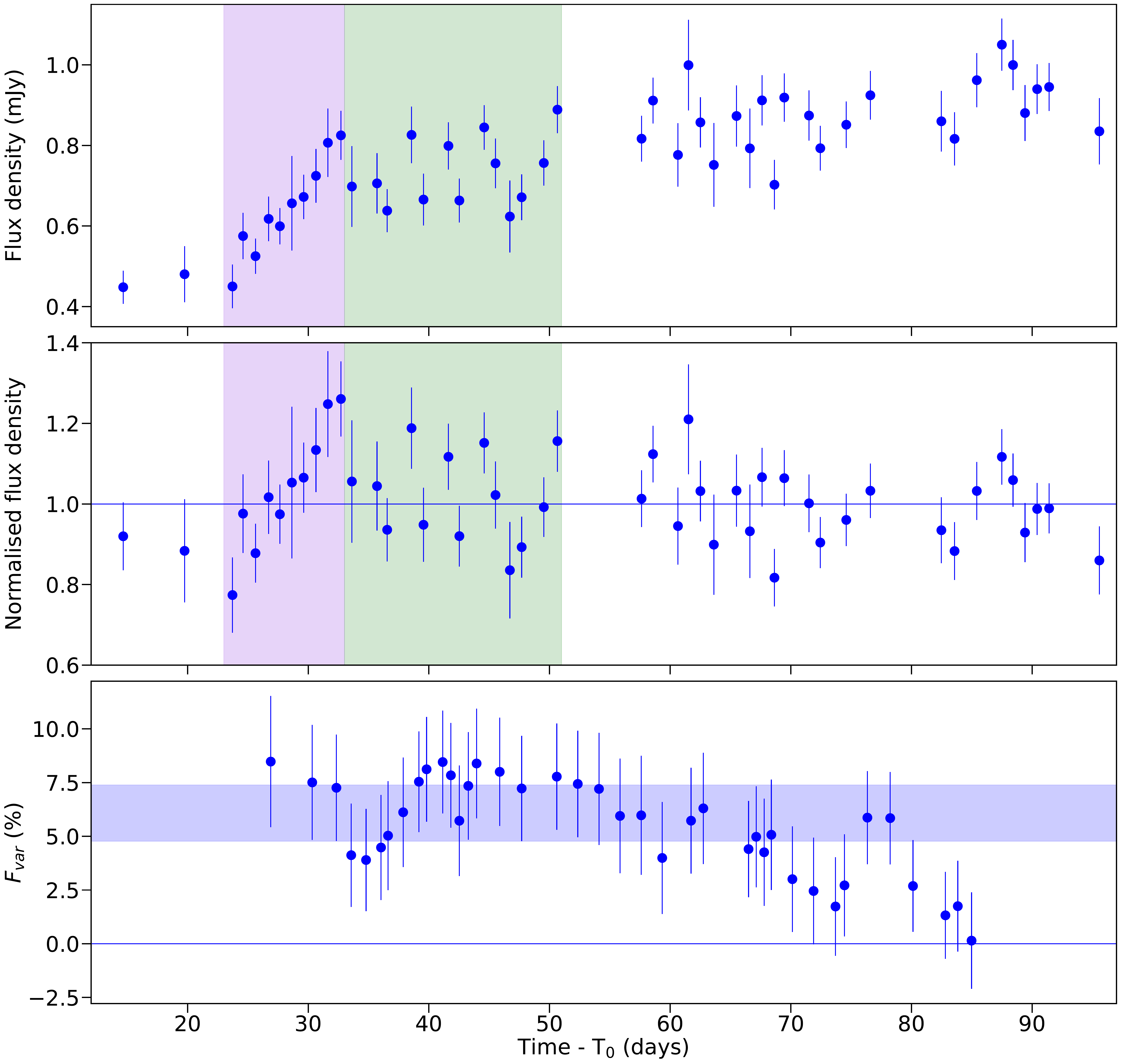}
    \caption{\textit{a}: The AMI--LA 15.5\,GHz light curve of AT2022cmc. The fractional variability is 16.5$\pm$1.3\%. \textit{b}: the 15.5\,GHz light curve normalised by dividing through by a single power law fit to the data. The fractional variability of the whole normalised light curve is 6.1$\pm$1.3\%. \textit{c}: the fractional variability of AT2022cmc at 15.5\,GHz as a function of time: each data point is the fractional variability of 15 days of observations plotted at the time average of those 15 days. The blue-shaded region is the time average fractional variability for the whole observing campaign. The purple and green shaded regions refer to the periods of ten-day and one-day variability.}
    \label{fig:cmc_var}
\end{figure*}

\subsubsection{AMI--LA}

Observations with AMI--LA \citep{amila, 2018MNRAS.475.5677H} started on 2022 February 26\textsuperscript{th} 00:40 UT (14.6\,days after the initial discovery). For each observation 3C 286 was used as both the interleaved phase calibrator and to set the absolute flux scale. We use the custom pipeline \textsc{reduce\_dc} (e.g. \citealt{reduce}) to flag instrumental issues, calibrate the bandpass response of the array, correct for atmospheric temperatures, and solve for phases on the interleaved calibrator which were then applied to the target field.
Because 3C 286 was also used as the complex gain calibrator, we reinitialized the sky-model for 3C 286 from the Perley-Butler 2017 standard \citep{perley2017} and  solved for complex gains (both amplitudes and phases) with a solution interval of 600\,s, deriving solutions for each of the eight frequency channels. We then image all observations for both fields using the \textsc{casa} task \textit{clean}, and measure the flux density of 3C 286 and AT2022cmc using \textit{imfit}. The resulting flux density was measured for 3C 286 is stable to better than 1.%..

\subsubsection{\textit{e}--MERLIN}

We obtained Rapid Response Time observations (PI: Rhodes, RR13002) of AT2022cmc with \textit{e}--MERLIN beginning with an initial epoch at 5\,GHz on 2022 February 27\textsuperscript{th} 02:30 UT (15.7\,days) with follow-up observations every three weeks. \textit{e}--MERLIN data are reduced using a \textsc{casa}-based (Version 5.8.0) pipeline \citep{2021ascl.soft09006M}. The pipeline averages, flags for radio frequency interference, calibrates and images the data. We do not detect the 5\,GHz in the first two epochs. From 40\,days onwards, a point source at around 100$\mu$Jy was persistently detected.

%The data are flagged for a priori excision of known areas of RFI, band edges and quacking. The dataset is averaged down to 4\,second integrations and 4 spectral windows each 128\,MHz wide. Bandpass (using J1407+2827) and initial gain (1329+3154 was used as a gain calibrator) calibration is performed. Flux scaling bootstraps the flux density from 3C 286 to all sources. Final band pass and gain calibration is performed and applied to the target field. The target field is split out and further flags are applied. Interactive cleaning and deconvolution is also performed in \textsc{casa} using the task \textit{tclean}. Table \ref{tab:merlin_obs} show the observation times, flux densities and upper limits.

\subsubsection{MeerKAT}
Observations with MeerKAT were awarded through an open-time call for proposals (PI: Rhodes, MKT-20185). Three observations were made over the first 100 days, each at 1.28\,GHz with a bandwidth of 0.856\,GHz.
%We obtained three epochs with MeerKAT through an open-time call proposal . Each observation was performed at % and lasted between five and seven hours: 40\,minute scans of the source were interleaved with 2\,minute scans of the phase calibrator: 3C 286. Each epoch was preceded by an additional 10\,minute scan of 3C 286 which also acted as the band-pass and flux calibrator. 
MeerKAT data are reduced using \textsc{oxkat}, a series of semi-automated \textsc{python} scripts \citep{oxkat}. The scripts flag and calibrate the data using standard procedures in \textsc{casa} \citep{CASA} then images are made using \textsc{wsclean} \citep{offringa-wsclean-2014}. A round of phase-only self-calibration is also performed. %The data are averaged down into 8\,second intervals and 1024 channels. The calibrator fields are flagged for RFI. Flags were applied to known bad channels as well as the first and last 100 channels. The calibrator fields are split out in order to calculate a spectral model from the primary to apply to the phase calibrator. Flux, gain and delay calibration tables are derived and applied to the target field. The flagging and calibration is performed using.
We do not detect any radio emission at the position of AT2022cmc in any of the three observations reported here.%, the observation times and upper limits for each epoch are reported in Table \ref{tab:MeerKAT_obs}. 

\subsection{Variability and brightness temperature calculations}

To interpret the observations we have made of AT2022cmc, we calculate the brightness temperature using different variability timescales. Here, we present the fractional root mean square (rms) variability which is required to check whether any variability observed is statistically significant using the method from \citet{2003MNRAS.345.1271V}. Then we demonstrate how to use a given variability timescale to calculate the brightness temperature and from there the relativistic Doppler factor.

The fractional rms variability examines the variability that is in excess of the contribution from the uncertainties associated with the measured flux densities. The fractional variability is given by 
\begin{equation}
F_{\mathrm{var}}=\sqrt{\frac{S^{2}-\overline{\sigma_{\mathrm{err}}^{2}}}{\bar{S_{\nu}}^{2}}}
\label{eq:var}
\end{equation}
where the variance is 

\begin{equation}
S^{2}=\frac{1}{N-1} \sum_{i=1}^{N}\left(S_{\nu,i}-\bar{S_{\nu}}\right)^{2}
\end{equation}
and the mean error is,
\begin{equation}
\overline{\sigma_{\mathrm{err}}^{2}}=\frac{1}{N} \sum_{i=1}^{N} \sigma_{\mathrm{err}, i}^{2} \text {. }
\end{equation} 
where $\bar{S_{\nu}}$ is the mean flux density and $N$ is the number of data points. The uncertainty associated with the fractional variability is given by:
\begin{equation}
\operatorname{err}\left(F_{\mathrm{var}}\right)=\sqrt{\left\{\sqrt{\frac{1}{2 N}} \cdot \frac{\overline{\sigma_{\mathrm{err}}^{2}}}{\bar{x}^{2}F_{\mathrm{var}}}\right\}^{2}+\left\{\sqrt{\frac{\overline{\sigma_{\mathrm{err}}^{2}}}{N}} \cdot \frac{1}{\bar{x}}\right\}^{2}}
\label{eq:var_err}
\end{equation}

Equations \ref{eq:var} and \ref{eq:var_err} are used to calculate whether the variability observed in the light curve is real and statistically significant. We consider any F\textsubscript{var} value greater than three times its associated uncertainty, $\operatorname{err}\left(F_{\mathrm{var}}\right)$ as statistically significant.

If the observed variability is real, we can calculate the brightness temperature of a radio source starting with:

\begin{equation}
T_B = \frac{S_{\nu} c^2}{2 k_B \Omega \nu^2}
\label{eq:TB}
\end{equation}

\noindent where $\Omega = R^2/D_{A}^2$, \textit{c} is the speed of light,\textit{ k\textsubscript{B}} is the Boltzmann constant, $\theta$ is the projected size of the source on the sky, $\nu$ is the observing frequency, $D_{A}$ is the angular diameter distance (i.e. the comoving distance: $D/(1+z)$ or the luminosity distance $D_L/(1+z)^2$) and $R$ is the radius of the source \citep{2011hea..book.....L}. All of the above variables are observable except for the radius ($R$), which can be inferred from an observed variability timescale: $R = c\Delta t_{\textrm{var}}/(1+z)$. Substituting these values into Equation \ref{eq:TB} gives \citep{1995ARA&A..33..163W}:

\begin{equation}
T_B = \frac{S_{\nu} D_{L}^2}{2 k_B \nu^2 t_{\textrm{var}}^2 (1+z)^2} 
\label{eq:TB_2}
\end{equation}

Brightness temperatures above 10\textsuperscript{12}\,K are not possible from synchrotron radiation \citep{1969ApJ...155L..71K}. Above 10\textsuperscript{12}\,K, the emitting region undergoes significant Compton cooling so that the brightness temperature drops back below 10\textsuperscript{12}\,K, called the inverse-Compton catastrophe. Some incoherent transients have brightness temperatures above 10\textsuperscript{12}\,K, such as gamma-ray bursts, where the cause of the high brightness temperatures is most likely relativistic beaming. Therefore, any brightness temperature measurements above 10\textsuperscript{12}\,K must originate from radiation that is strongly beamed into our line of sight.

\noindent We use the brightness temperature measurements to infer the relativistic Doppler factor by substituting the following into Equation \ref{eq:TB_2}: $S_{\nu} = I_{\nu}\Omega$, $\nu = \nu'/(1+z)$, the angular radius $\theta = (\delta_{\textrm{var}}c\Delta t_{\textrm{var}})/(D_{A} (1+z))$, and I\textsubscript{$\nu$}/$\nu^{3}$ = I$'$\textsubscript{$\nu$}/$\nu'^{3}$. Equation \ref{eq:TB_2} can be rearranged to obtain the Doppler factor in terms of the observed brightness temperature (T\textsubscript{var}) and a given rest frame temperature ($\textrm{T}'_{\textrm{var}}$):

\begin{equation}
\delta_{\textrm{var}} = \sqrt[3]{(1+z)\frac{T_{\textrm{var}}}{T_{\textrm{var}}'}} = \frac{1}{\Gamma(1-\beta\cos{\phi})}
\label{eq:df}
\end{equation}

\noindent where $\Gamma$ is the bulk Lorentz factor, $\beta$ is the velocity of the outflow material as a fraction of the speed of light and $\phi$ is the angle of the outflow to the line of sight. A result of Doppler boosting is that radiation is beamed into a cone within an opening angle, $\phi \approx 1/\Gamma$. For the work presented in this paper, we assume a rest frame brightness temperature of T$'$\textsubscript{var} = 10\textsuperscript{12}\,K. However, it is possible that the rest frame temperature is considerably lower \citep[e.g.][]{1994ApJ...426...51R, 1999ApJ...511..112L}. From Equation \ref{eq:df}, one can see that by decreasing T$'$\textsubscript{var}, the Doppler factor we infer would increase. Therefore, the Doppler factors we present in this work are lower limits.

\section{Results}\label{results}

Figure \ref{fig:lc} shows radio light curves from MeerKAT \citep[1.3\,GHz][]{2016mks..confE...1J}, \textit{e}--MERLIN (5\,GHz) and AMI--LA (15.5\,GHz) along with data points at 5, 8.5, 10.5, 15.9 and 33.5\,GHz from the Karl G. Jansky Very Large Array \citep{2022arXiv221116530A}. In all bands (except 1.3\,GHz, where AT2022cmc was not detected), the light curves show a slow rise over the duration of the respective observing campaigns. A radio counterpart was first detected with the AMI--LA from 14\,days post-discovery \citep{2022GCN.31667....1S} and at 5\,GHz with \textit{e}--MERLIN at 40 days post-discovery. The high time cadence of the 15.5\,GHz AMI--LA light curve shows day-timescale variability with an underlying slowly rising light curve.

To parameterize and understand the short timescale variations occurring within the radio emitting region as observed at 15.5\,GHz we use the fractional variability. The fractional variability examines the variability that is in excess of the contribution from the uncertainties associated with the measured flux densities \citep{2003MNRAS.345.1271V}. 

We measure a fractional variability of 16.5$\pm$1.3\,\% for the whole 15.5\,GHz data set which considers both the underlying long-term flux density increase as well as the day-to-day variability. In order to determine if the short timescale variability is real, we fit a power law to the 15.5\,GHz light curve (shown in Figure \ref{fig:cmc_var}a) and divide the light curve data by the fit (Figure \ref{fig:cmc_var}b) and obtain an excess variability of AT2022cmc is 6.1$\pm$1.3\,\%.

The observed variability is unlikely to be a result of scintillation at 15.5\,GHz. For the position of AT2022cmc on the sky, the AMI--LA observing band is firmly in the weak scintillation regime \citep{2002astro.ph..7156C}. Any effects due to weak scintillation would cause variability on a timescale of approximately 1\,hour with an amplitude of about 30\% \citep{1997NewA....2..449G}. We measure no statistically significant variability on this timescale. %We measure statistically insignificant intra-observation variability on the order of 30\% in the first two epochs but it cannot reproduce the observed epoch-to-epoch variability. 

Given that it is unlikely that scintillation is the origin of the observed variability, it is possible that it originates from the TDE. A significant contribution ($\sim$ one third) to the fractional variability measurement originates from the data points between 23 and 33\,days post-discovery where the flux density increases by 50\% over 10\,days (the highlighted purple region in Figure \ref{fig:cmc_var}). A variability timescale ($\Delta t$) of 10\,days corresponds to a very small emission region ($c \Delta t/(1+z)$) of $\lesssim$ 1.2$\times$10\textsuperscript{16}\,cm and a high brightness temperature of (2.0$\pm$0.4)$\times$10\textsuperscript{13}\,K.% Incoherent (e.g. synchrotron) sources with brightness temperatures above 10\textsuperscript{12}\,K undergo rapid Compton cooling bringing the temperatures back down below 10\textsuperscript{12}\,K \citep{2007A&A...463..145T}. Therefore, a measurement of an apparent brightness temperature above 10\textsuperscript{12}\,K from an incoherently emitting source is most likely to be the result of Doppler boosting.

\begin{figure}
    \centering
    \includegraphics[width = \columnwidth]{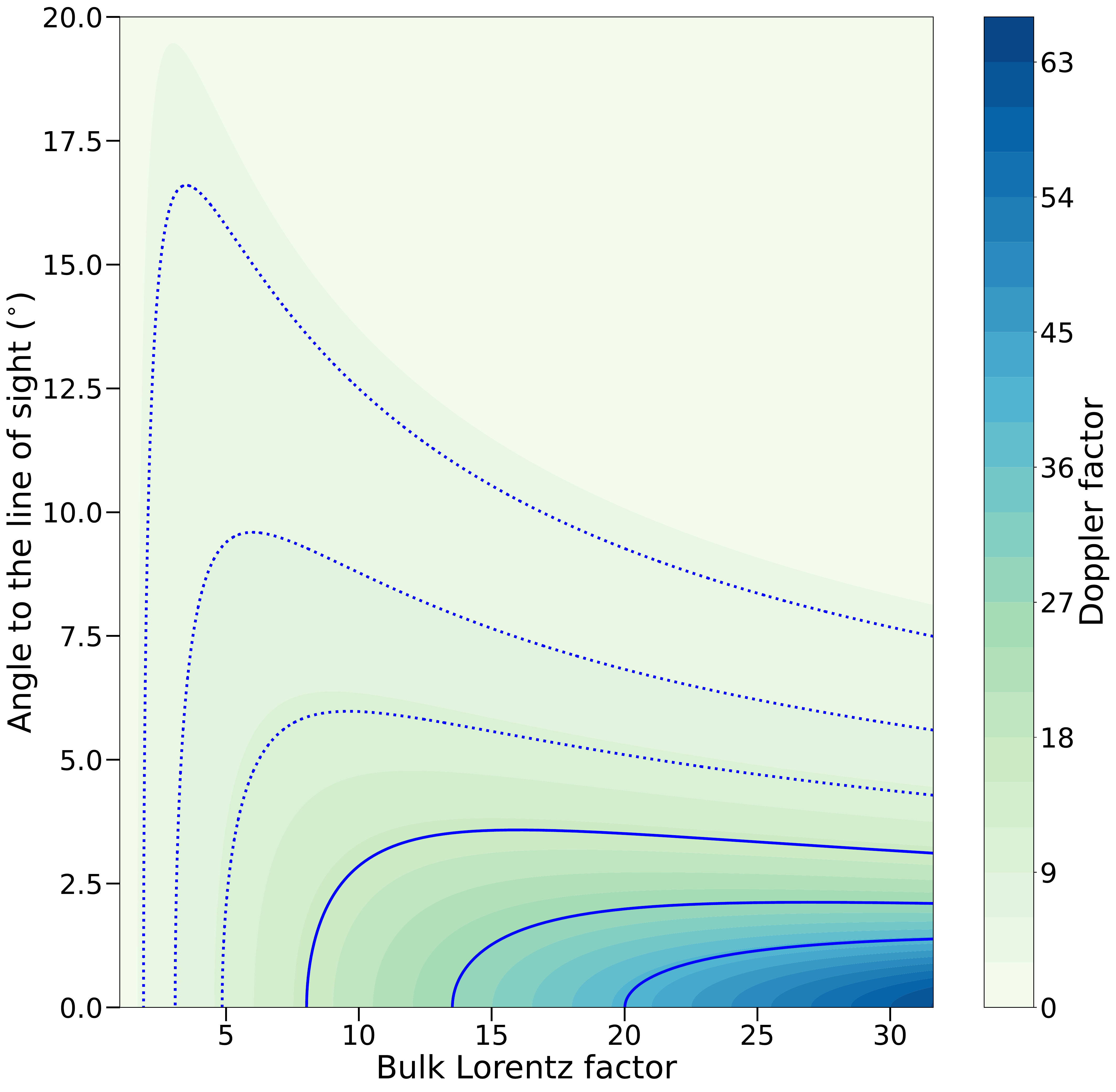}
    \caption{A contour plot showing the relativistic Doppler factor ($\delta_{\textrm{var}}$) as a function of angle to the line of sight and bulk Lorentz factor. The solid blue lines correspond to Doppler factors calculated from the day-timescale variability using rest frame brightness temperatures of 10\textsuperscript{12}, 2$\times$10\textsuperscript{11} and 5$\times$10\textsuperscript{10}\,K. The dotted lines are the Doppler factors from the variability timescale of 10 days using the same three rest frame brightness temperatures.}
    \label{fig:dopp}
\end{figure}

From the brightness temperature of (2.0$\pm$0.4)$\times$10\textsuperscript{13}\,K, we use Equation \ref{eq:df}, and $\textrm{T}_{\textrm{var}}'$ = 10\textsuperscript{12}\,K,  to derive a Doppler factor of 4. There are two other rest frame brightness temperature limits that are often used in the literature: 2$\times$10\textsuperscript{11}\,K \citep{1994ApJ...430..550S} and 5$\times$10\textsuperscript{10}\,K \citep{1994ApJ...426...51R}. By using these other values in our calculations, we obtain Doppler factors of 6 and 10, respectively. The dotted blue lines in Figure \ref{fig:dopp} shows the allowed values of $\Gamma$ and $\phi$ for three values of $\delta_{\textrm{var}}$ quoted above. 
%If we were to use a lower rest frame brightness temperature, the lower limit we place on the Doppler factor would increase further. 

%I note that there are two other possible lower limits, the claimed self-absorption lower limit of 2e11 (Krolik) or the 'equipartition' lower limit of a few e10 (Readhead). Should we mention these explicitly?

Between days 33 and 51 post-discovery (the green highlighted region in Figure \ref{fig:cmc_var}), the AMI--LA light curve shows evidence of day-timescale variability. Shorter timescale variability corresponds to an even higher brightness temperature of (2.0$\pm$0.4)$\times$10\textsuperscript{15}\,K. We calculate values of $\delta_{\textrm{var}}$ to be 16, 27 and 30, for T$'$\textsubscript{var} = 10\textsuperscript{12}, 2$\times$10\textsuperscript{11}\,K and 5$\times$10\textsuperscript{10}\,K, respectively \citep{1994ApJ...430..550S, 1994ApJ...426...51R}. All three values are marked by solid blue lines on Figure \ref{fig:dopp}.

Given the consistently high cadence data over the first 100\,days after the TDE discovery, we can also search for changes in variability amplitude. Figure \ref{fig:cmc_var}(c) shows the percentage fractional variability as a function of time using bins of 15 days. There is significant variability for the first 70 days after which the epoch-to-epoch variability ceases. There is no evidence in the data to link the reduction in variability to any spectral index variations. For the first 100\,days, we measure a spectral index of 1.9$\pm$0.1, consistent with what is expected from self-absorbed synchrotron emission.

\section{Discussion}\label{discussion}

Using model-independent analysis, from the Doppler factor calculation (Equation \ref{eq:df}), we infer from day-timescale variability, that the bulk Lorentz factor of the outflowing material $\gtrsim$8 for a jet that is pointing directly towards Earth, i.e. the angle to the line of sight is zero degrees. \citet{2022NatAs.tmp..252P} also inferred a slightly higher bulk Lorentz factor through SED modelling. We also find that after the first 70\,days post-discovery that there is a reduction in variability. This decrease could reflect a reduction in mass accretion rate \citep{1994ApJ...430L..93R, 2013MNRAS.429L..20M}, or an intrinsic reduction of variability at the shock front.

In order to produce such a highly relativistic outflow, vast amounts of kinetic energy are required, so much so that the observed emission cannot originate from a spherical outflow. \citet{2022arXiv221116530A} used the X-ray emission to estimate an isotropic equivalent kinetic energy of $\approx10^{53-54}$\,erg. For an estimate of $10^{54}$\,erg, at least 20\,M\textsubscript{\(\odot\)} of material is required (by assuming an optimistic efficiency of 10\% and considering that 50\% of the material will not be accreted on to the SMBH) to produce the observed emission if the outflow is isotropic. This problem is alleviated if instead the outflow is collimated into a jet-like outflow and beamed into our line of sight.

Evidence of relativistic outflows has been found in many classes of systems, both galactic and extragalactic. Using Very Long Baseline Interferometry \citep[e.g. ][]{1981Natur.290..365P, 1985ApJ...289..109U, 2001ApJS..134..181J} and multi-wavelength monitoring programs \citep[e.g.][]{2017MNRAS.466.4625L}, Doppler factors as high as those shown in Figure \ref{fig:dopp} have been found in blazars. Gamma-ray bursts, some of the most powerful explosions known, also require high bulk Lorentz factors of at least 100 \citep{2010MNRAS.402.1854Z, 2018A&A...609A.112G}. High angular resolution observations in multiple bands support the requirement for high launch Lorentz factors where gamma-ray burst jets have Lorentz factors of around 5, at tens to hundreds of days post-burst \citep{2004ApJ...609L...1T, 2022Natur.610..273M}. We note that, in the case of gamma-ray bursts, it is often assumed that the angle to the line of sight is zero and Lorentz factors are quoted instead of Doppler factors.

Within the Milky Way, X-ray binaries have a larger range of Doppler factors. The average jet angle to the line of sight is around 60\textsuperscript{$\circ$} meaning that in many cases the emission we observe is actually deboosted. When Lorentz factors are calculated considering a given systems inclination angle, they are much lower than those measured for extra-galactic systems, with values around 2 \citep{2020NatAs...4..697B, 2022MNRAS.511.4826C}.

The Doppler factor we infer from the ten-day-timescale variability observed in AT2022cmc are consistent with the highest superluminal velocities measured in blazar systems \citep{2005AJ....130.1418J, 2009A&A...494..527H, 2017MNRAS.466.4625L} as well as gamma-ray burst jets. The high Lorentz factor measured for AT2022cmc most likely arises from the transient nature of AT2022cmc, the short injection energy resulted in a higher Doppler factor for a short period of time followed by the deceleration of the jet.

%Evidence of relativistic outflows are commonly found in blazars using very long baseline interferometry \citep[e.g. ][]{2001ApJS..134..181J} and multi-wavelength monitoring programs \citep[e.g.][]{2017MNRAS.466.4625L}. The bulk Lorentz factors we infer from the day-timescale variability are only consistent with blazar observations with the highest superluminal velocities \citep{2005AJ....130.1418J, 2009A&A...494..527H, 2017MNRAS.466.4625L}. The high $\Gamma$ measured for AT2022cmc most likely arises from the transient nature of AT2022cmc, the short injection energy resulted in a higher Doppler factor for a short period of time followed the deceleration of the jet. 

\begin{figure}
     \centering
         \includegraphics[width=\columnwidth]{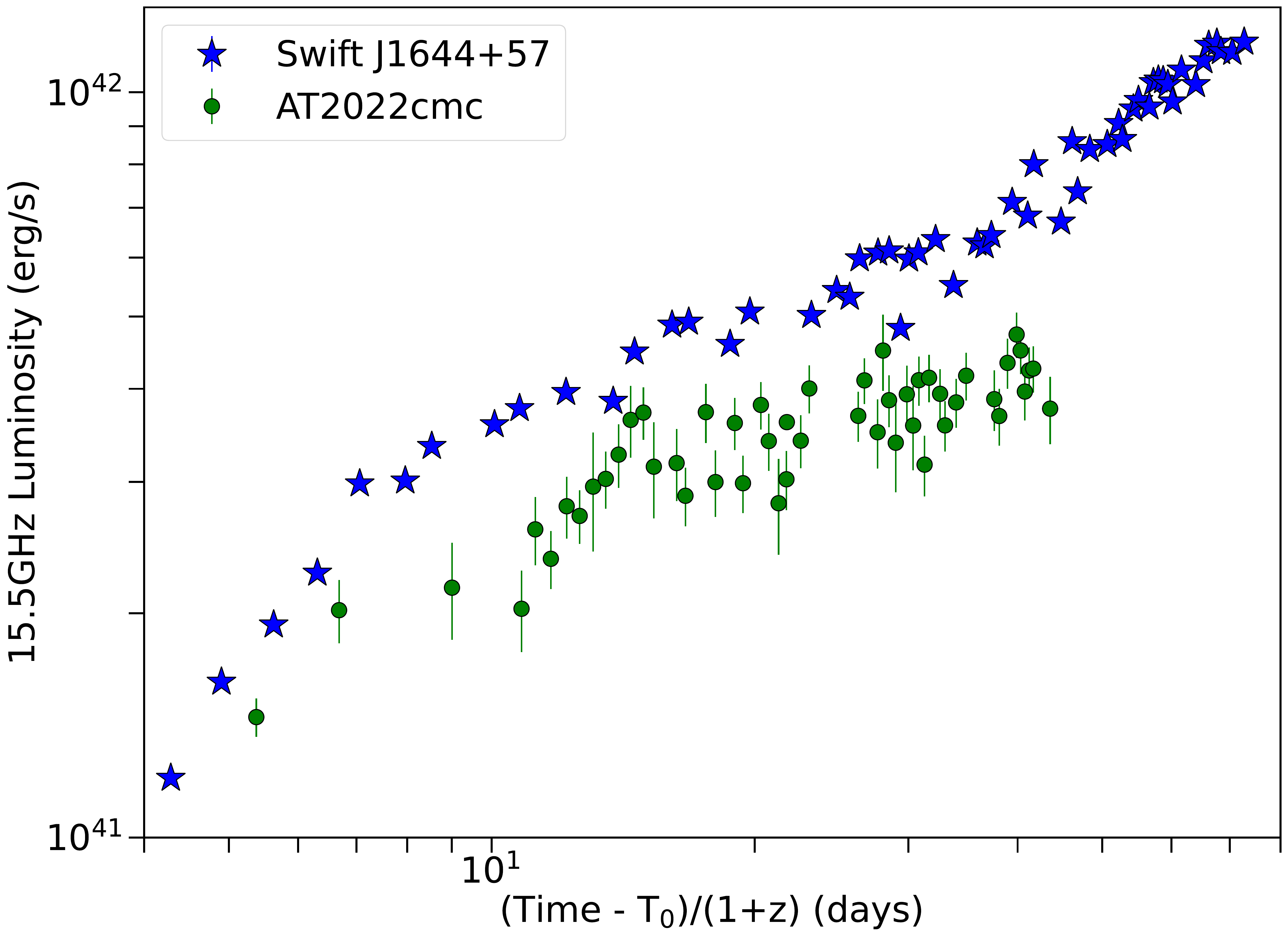}
        \caption{The 15.5\,GHz luminosities of \textit{Swift} J1644 (blue stars) and AT2022cmc (green circles) for the first 100 days in the rest frame of the respective events. The two events show a similar long term evolution for the period in which we have overlap i.e. between 6 and 45 rest frame days. We do not consider the Doppler boosting in calculating the luminosities. }
        \label{fig:luminosities}
\end{figure}

Comparison of the 15.5\,GHz luminosities of \textit{Swift} J1644+57 and AT2022cmc over the first 100\,days corrected for their respective redshifts since their respective discovery days (Figure \ref{fig:luminosities}).
%Finally, Figure \ref{fig:luminosities} shows the 15.5\,GHz luminosity evolution of \textit{Swift} J1644+57 (blue stars) and AT2022cmc (green circles) for the first 100\,days after their respective discoveries, corrected for their respective redshifts. 
The evolution of both \textit{Swift} J1644+57 and AT2022cmc show remarkable similarities over the first 40 days in their respective rest frames, where the light curves follow a power law rise of approximately $S_{\nu} \propto t^{0.4}$ \citep{2012ApJ...748...36B}. Unlike in AT2022cmc, \textit{Swift} J1644+57 shows no statistically significant variability. To infer the same bulk Lorentz factor in the AMI--LA data of \textit{Swift} J1644+57 as we have for AT2022cmc we would have to observe variability on a timescale of less than one day, a timescale not sampled by the radio follow-up campaigns.

\section{Conclusions}

Our radio campaign to observe the newly discovered relativistic TDE AT2022cmc has produced high cadence multi-frequency data set spanning the first 100 days post-discovery. The 15.5\,GHz light curve shows short timescale variability which corresponds to a very high brightness temperature implying the presence of Doppler boosting and providing a model-independent confirmation of a relativistic outflow. The analysis we have performed here is vital in our long-term understanding, modelling and interpretation of AT2022cmc and other relativistic TDEs. This is the first direct evidence of Doppler beaming in any type of TDE and confirms that such systems are truly relativistic.

\section*{Acknowledgements}

L. R. acknowledges the support given by the Science and Technology Facilities Council through an STFC studentship. We thank the Mullard Radio Astronomy Observatory staff for scheduling and carrying out the AMI--LA observations. The AMI telescope is supported by the European Research Council under grant ERC-2012-StG-307215 LODESTONE, the UK Science and Technology Facilities Council, and the Universities of Cambridge and Oxford. The MeerKAT telescope is operated by the South African Radio Astronomy Observatory, which is a facility of the National Research Foundation, an agency of the Department of Science and Innovation. \textit{e}--MERLIN is a National Facility operated by the University of Manchester at Jodrell Bank Observatory on behalf of STFC, part of UK Research and Innovation. This research has made use of NASA’s Astrophysics Data System, and the Python packages \textsc{numpy} \citep{5725236} and \textsc{matplotlib} \citep{4160265}.

%%%%%%%%%%%%%%%%%%%%%%%%%%%%%%%%%%%%%%%%%%%%%%%%%%
\section*{Data Availability}

All the data used in this paper are available in the online Appendices along with AMI--LA and MeerKAT radio maps and variability consistency checks.



\bibliographystyle{mnras}
\bibliography{example} % if your bibtex file is called example.bib

% Don't change these lines
\bsp	% typesetting comment
\label{lastpage}
\end{document}